\documentclass[conference]{IEEEtran}
\IEEEoverridecommandlockouts

\usepackage{cite}
\usepackage{amsmath,amssymb,amsfonts}
\usepackage{algorithmic}
\usepackage{textcomp}
\usepackage{xcolor}
\usepackage{amsmath}
\usepackage{listings}
\usepackage{url}
\usepackage{booktabs}
\usepackage{multirow}

\usepackage[normalem]{ulem}
\useunder{\uline}{\ul}{}

\usepackage[pdftex]{graphicx}
\usepackage{subcaption}

\def\BibTeX{{\rm B\kern-.05em{\sc i\kern-.025em b}\kern-.08em
    T\kern-.1667em\lower.7ex\hbox{E}\kern-.125emX}}

\begin{document}

\title{AV1 Motion Vector Fidelity and Application for Efficient Optical Flow
\thanks{This work was funded by the Horizon CL4 2022, EU Project Emerald, 101119800.}
}

\author{\IEEEauthorblockN{Julien Zouein, Vibhoothi Vibhoothi, Anil Kokaram}
\IEEEauthorblockA{
Sigmedia Group, \\
Department of Electronic and Electrical Engineering, \textit{Trinity College Dublin}, Dublin, Ireland \\
\{zoueinj, vibhootv, anil.kokaram\}@tcd.ie
}
}

\maketitle

\begin{abstract}
This paper presents a comprehensive analysis of motion vectors extracted from AV1-encoded video streams and their application in accelerating optical flow estimation. We demonstrate that motion vectors from AV1 video codec can serve as a high-quality and computationally efficient substitute for traditional optical flow, a critical but often resource-intensive component in many computer vision pipelines. Our primary contributions are twofold. First, we provide a detailed comparison of motion vectors from both AV1 and HEVC against ground-truth optical flow, establishing their fidelity. In particular we show the impact of encoder settings on motion estimation fidelity and make recommendations about the optimal settings. Second, we show that using these extracted AV1 motion vectors as a ``warm-start'' for a state-of-the-art deep learning-based optical flow method, RAFT, significantly reduces the time to convergence while achieving comparable accuracy. Specifically, we observe a four-fold speedup in computation time with only a minor trade-off in end-point error. These findings underscore the potential of reusing motion vectors from compressed video as a practical and efficient method for a wide range of motion-aware computer vision applications.
\end{abstract}


\section{Introduction}

The estimation of motion between video frames is a fundamental task that bridges the fields of video compression and computer vision. In video compression, motion estimation is essential for exploiting temporal redundancy. In computer vision, the equivalent concept, optical flow, provides vital information for tasks ranging from object tracking to autonomous navigation. However, the computational cost of generating accurate optical flow often creates a significant bottleneck in real-world applications.

Given that the vast majority of digital video is stored and transmitted in a compressed format, the motion vectors required for decoding are readily available within the bitstream. This has led to research into reusing these pre-computed motion vectors as a computationally inexpensive proxy for optical flow \cite{kantorov}. Building on prior work that has primarily focused on older codecs like H.265, this paper investigates the potential of the modern, open-source AV1 codec for this purpose. The main contributions of this paper are:

i) A quantitative analysis of the motion vectors extracted from AV1 bitstreams, compared with those from HEVC and ground-truth optical flow, to assess their quality and suitability for computer vision tasks.

ii) An investigation into the impact of using these extracted sparse motion fields to accelerate a modern, deep learning-based optical flow algorithm, RAFT, by using the motion vectors as a "warm-start" to reduce computational overhead.

\section{Background}

The use of metadata from compressed video bitstreams, particularly motion vectors, for improving the efficiency of computer vision tasks has been an area of active research . The core idea is to leverage the motion information already computed by the video encoder, thus avoiding the computationally expensive process of recalculating it from raw pixel data.
In 2014, Kantarov et al.~\cite{kantorov} showed that motion vectors from the H.264 codec could be used for action recognition. By avoiding the analysis of raw pixels, they report an accelerated analysis time with a factor of 2 improvement over traditional feature generation. More recently, Ehrlich et al. have developed a deep learning architecture for compressed video quality enhancement~\cite{MetaBit} that leverages structural and motion information embedded in the bitstream to restore fine details in compressed videos. While the use of modern motion estimation techniques in video compression has yielded mixed results\cite{ringis2018mvreuse}, the idea of exploiting motion information in the bitstream to accelerate optical flow estimation dates back to about 2018~\cite{hevc-epic}.

Rüfenacht and Taubman proposed HEVC-EPIC~\cite{hevc-epic}, where they re-used EPICFlow~\cite{EpicFlow} framework replacing the feature matching part with Motion Vectors extracted from HEVC bitstream, to yield a fast optical flow estimate. In 2020 Taubman refined this approach~\cite{Fast-Optical-Flow-Extraction-From-Compressed-Video}, where they reformulate a slow optimization problem into a fast confidence-weighted filtering problem.

The use of Deep Learning techniques to refine Motion Vectors extracted from compressed bitstream has also been explored. The RAFT~\cite{raft} technique aims to estimate optical flow with high-quality control using an iterative process. In 2022, Zhou et al.~\cite{mvflow} have introduced a framework that uses motion vectors from the bitstream as input to a RAFT-like deep neural network~\cite{raft}, which iteratively refines the motion field to produce accurate optical flow. Their system is computationally heavier, using a combination of 3 Convolutional Neural Networks to pre-process motion information before attempting to warm-start their model.

Despite the progress in this area, previous work has not investigated the impact of encoder presets and configuration on the quality of the initial sparse motion field using HEVC and AV1. This has a direct impact on the fidelity and usefulness of their motion vectors. 
Our work addresses these gaps by systematically evaluating the quality of motion vectors produced by modern codecs and by carefully choosing the encoder parameterisation for high-quality, computationally efficient optical flow estimation.

\section{Extraction and pre-processing of the Motion Vectors}

Hybrid video codecs like AV1 and HEVC share a block-based prediction architecture.
AV1 supports a wider range of block sizes (4×4 to 128×128) compared to HEVC (4×4 to 64×64). 
Each block in a frame can be associated with a motion compensation vector indicating its source location from an already encoded frame (Reference Frame). AV1 supports a total of seven reference frames~\cite{av1:adaptivepred, zhao2021tool}, which can be set up based on coding configuration to be forward or backwards. 

The key observation is that the raw motion from the bitstream is sparse and unevenly sampled. We therefore need to create a dense, evenly sampled field for further refinement.

\subsection{Motion Vector Extraction and Optic Flow Initialisation}
We first manipulate the raw (and sparse) motion vectors extracted from the bitstream to create an initial flow field at the pixel resolution. That flow field must contain vectors  that reference the previous frame only. 

{\noindent \textbf{Reference Frame Normalisation.}}
Motion vectors (MV) within a single frame $n$ can point to different reference frames (e.g., the previous frame $n-1$, or one or several frames in the past $n-k$).
To create a coherent motion field relative to the immediately preceding frame $n-1$, we need to normalise each vector. Given the motion vector $v_{n,m}(p, q)$ at a block $(p,q)$ in frame $n$, we determine reference frame $m$ by accessing
the \texttt{order\_hint} syntax element metadata in  the bitstream. This maps the reference frame type (e.g., LAST1, LAST2, GOLDEN, ALTREF, etc) to an absolute frame index. Note that \texttt{order\_hint} is cyclic, which wraps after 128 frame indexes, this significantly complicates finding $m$. 

Given the motion vector $v_{n,m}(p, q)$  we simply determine the {\em normalised} vector $v_{n,n-1}(p,q)$ as
\begin{equation}
    \begin{aligned}
        v_{n,n-1}(p,q) = \frac{v_{n,m}(p,q)}{|n-m|}
    \end{aligned}
\end{equation}

{\noindent \textbf{Handling Missing Motion Data.}}
Blocks encoded using INTRA prediction or designated as SKIPPED do not have associated motion vectors, resulting in $(0.0)$ motion vectors. In some cases the motion in the opposite direction can have non-zero motion. In that case we
 use Bidirectional Motion Vector Completion (BMVC~\cite{bmvc}), to infer the missing motion from its corresponding opposite motion vector. 
Hence the missing motion $v_{n,n-1}(p,q)$, for example may be inferred from $v_{n,n+1}(p,q) = -v_{n,n-1}(p,q)$. If there is no {\em opposite} motion then we leave the cell at zero.

{\noindent \textbf{Upsampling Motion Field data to Frame Resolution.}}
The motion field in the bitstream is processed at the block level, with the smallest unit a 4$\times$4 pixel block. 
To generate a dense field matching the original frame resolution, we upsample using a zero-order hold, replicating each vector across all pixels within its corresponding 4$\times$4 block. To overcome padding issues, we apply appropriate cropping to the upsampled motion field to ensure dimensions align with the original video resolution.

\begin{figure*}
    \centering
    \begin{tabular}{cccc}
      \includegraphics[width=0.22\linewidth]{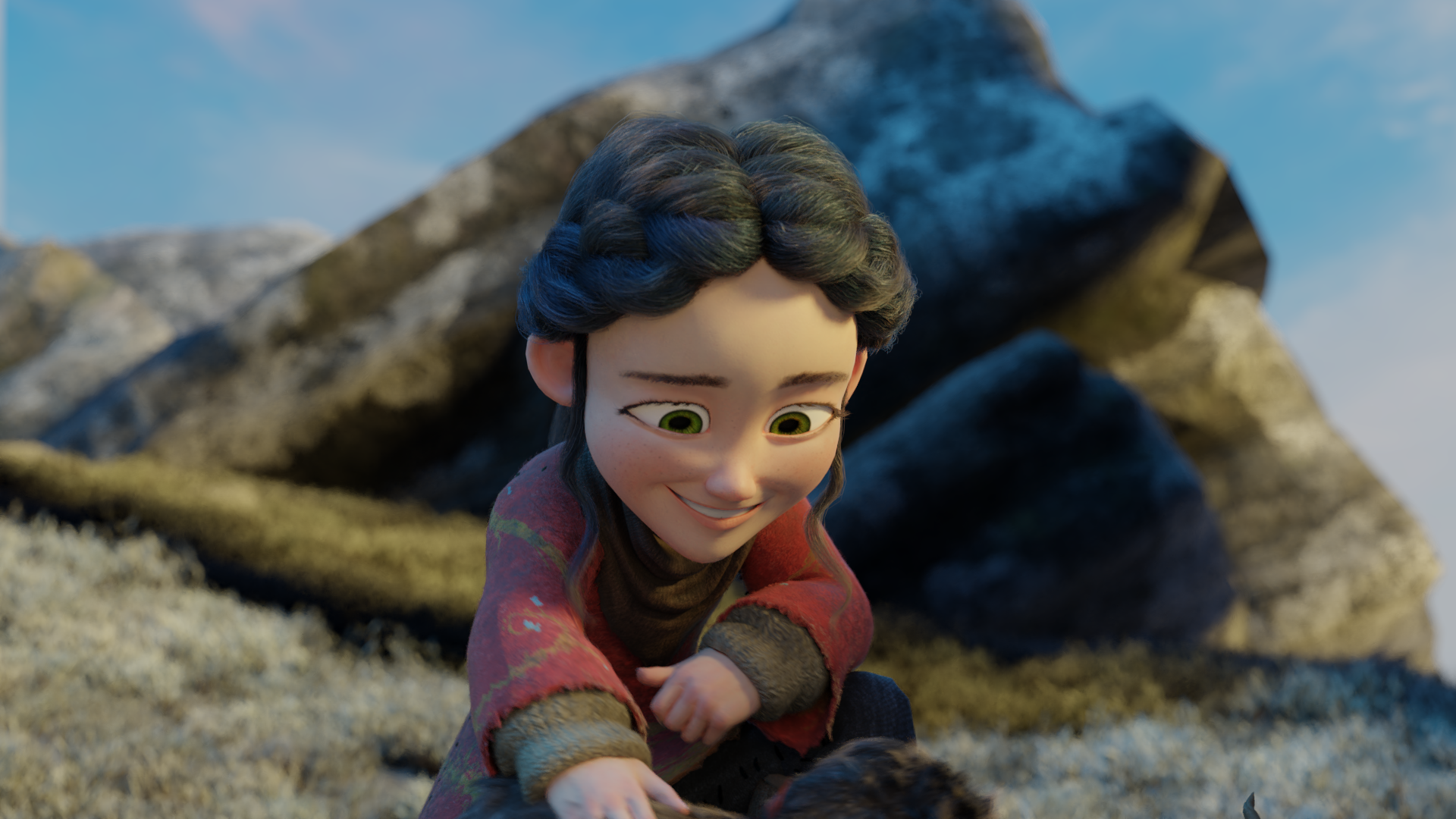}   & 
      \includegraphics[width=0.22\linewidth]{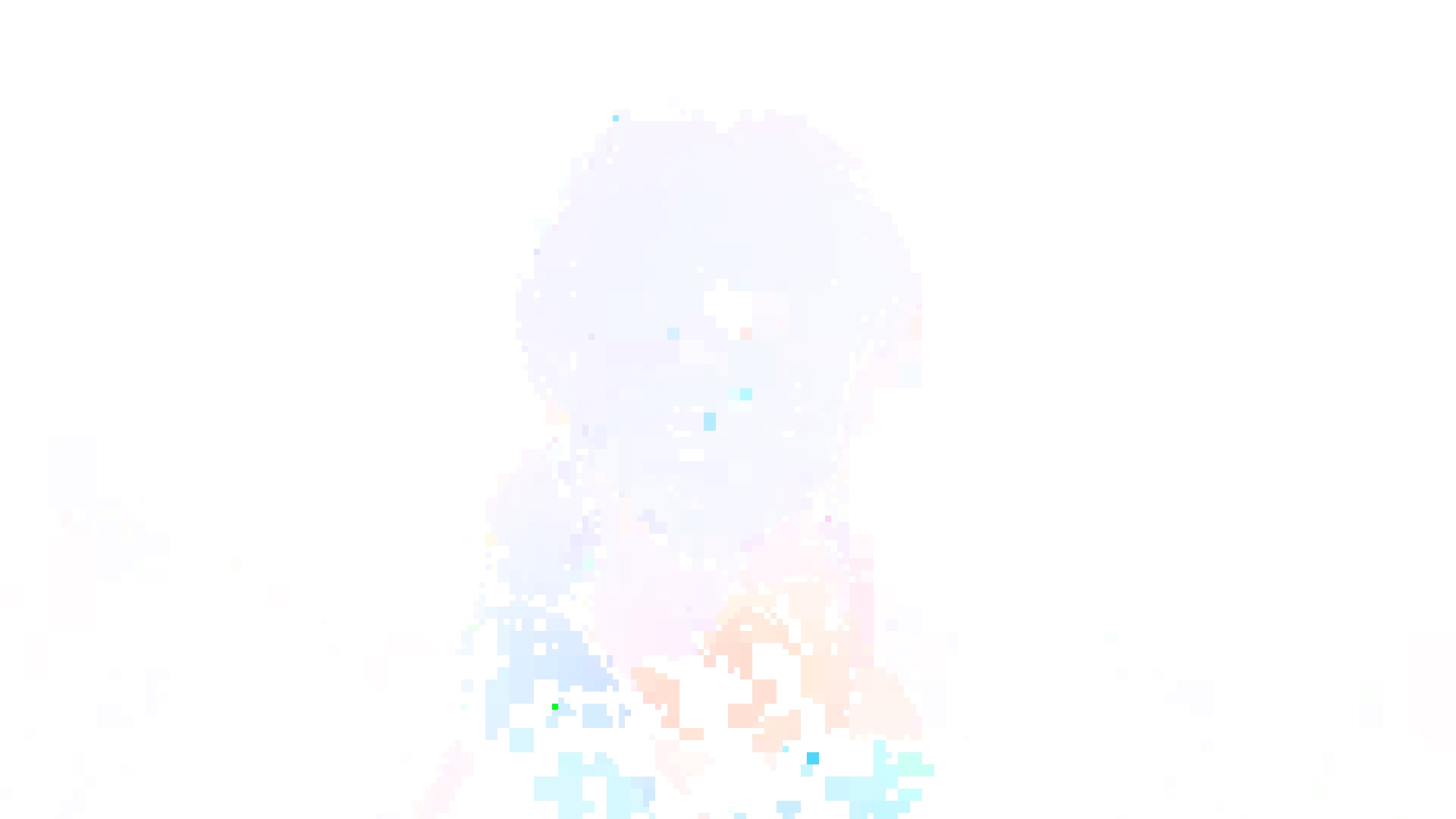} & 
      \includegraphics[width=0.22\linewidth]{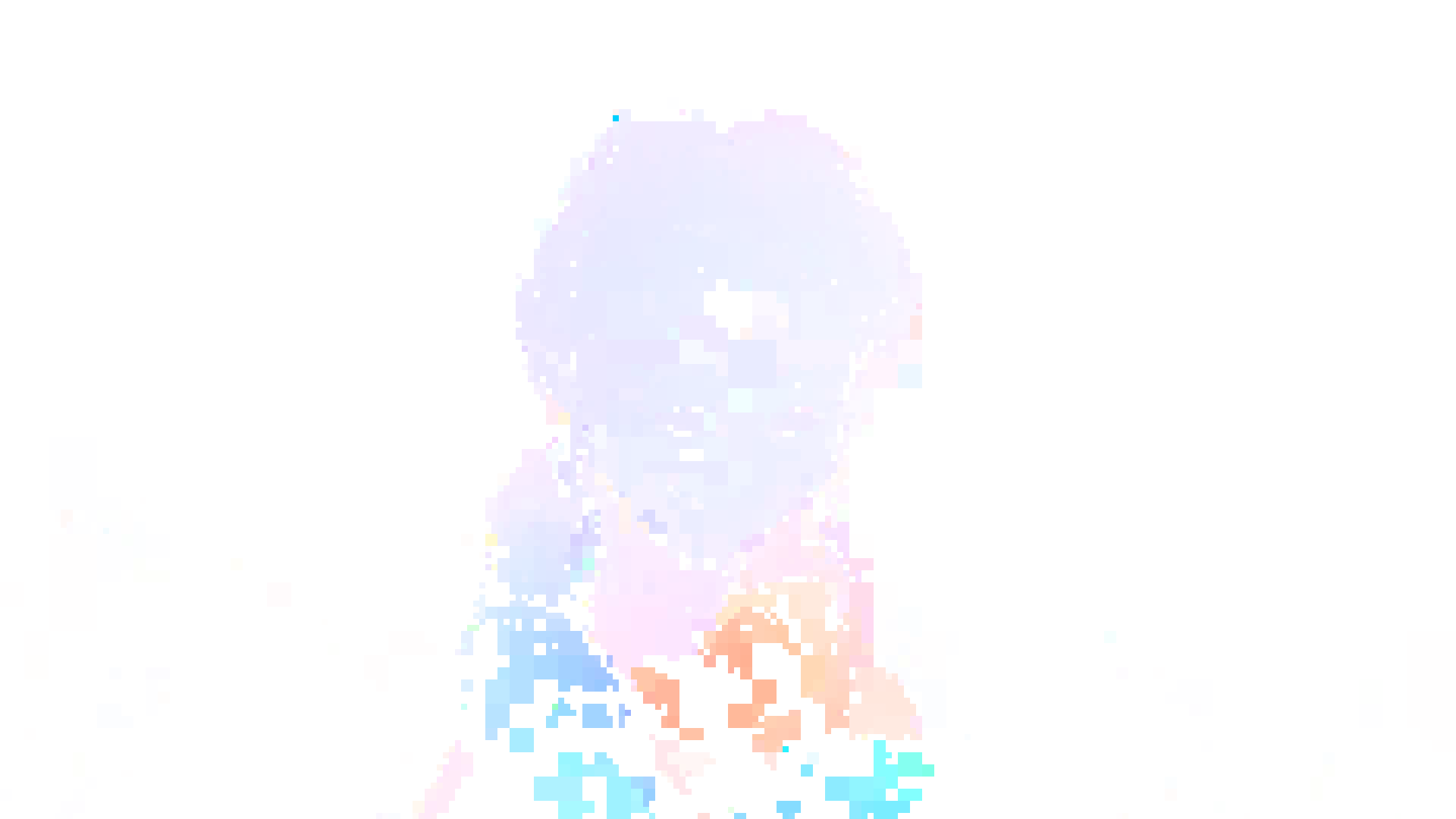} &
      \includegraphics[width=0.22\linewidth]{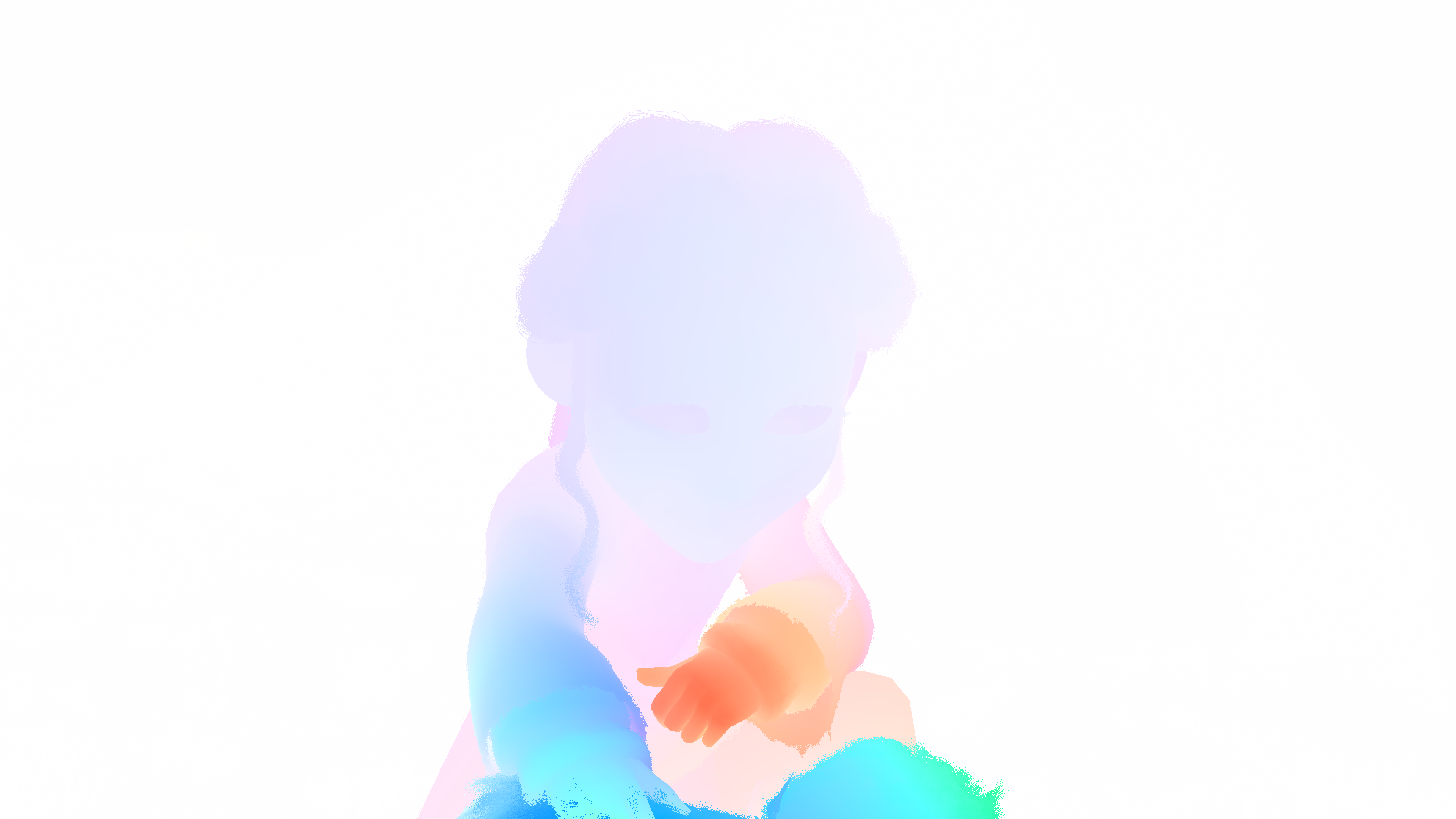} 
    \end{tabular}
    \caption{Visualisation of the processing of the extracted motion. Left to Right: RGB frame, Motion Field extracted from AV1, Processed Motion Field from AV1, Reference Optical Flow given by Spring Dataset.}
    \label{motion-extraction-and-processing}
\end{figure*}

Figure~\ref{motion-extraction-and-processing} illustrates the motion field at each stage of processing. The sparse motion field is reasonably populated by the end of the pipeline and already has some semblance of the actual ground truth flow field.

\subsection{Refinement of motion vectors}

The normalised upsampled motion field, $v_{n-1}$, exhibits blocking artefacts and lacks pixel-level smoothness. 
We improve this situation by employing RAFT (Recurrent All-Pairs Field Transforms~\cite{raft}), a state-of-the-art deep learning model for optical flow, as a refinement method. 
Although not originally designed for this task, RAFT's architecture supports a ``warm-start'' initialisation, where an existing motion field is used as a starting point for its iterative refinement process. 

RAFT is trained in 4 stages. The first stage uses FlyingChair dataset~\cite{flyingchairs} for training, giving ``RAFT-chairs'' weights. Second stage re-uses ``RAFT-chairs' to train on the FlyingThings dataset~\cite{flyingchairs} giving ``RAFT-things'' weights. Third stage trains on a combination of MPI Sintel dataset~\cite{mpi-sintel}, FlyingThings dataset, HD1K dataset~\cite{hd1k} and KITTI dataset~\cite{kitti}, generating ``RAFT-sintel''. Finally the last stage  focuses on KITTI dataset generating ``RAFT-kitti''.
We use RAFT after the final training stage for our first test, denoted as {\em RAFT} in Figure~\ref{RAFT-performances}.

RAFT was never trained using ``warm-start'', especially not with a blocky motion field extracted from bitstreams. To test the impact of training RAFT using ``warm-start'' we added an extra-stage, reusing the weights ``RAFT-kitti'' and training using MPI-Sintel dataset. We build two fine tuned models, one without any ``warm-start'': RAFT+;  and one using warm start with AV1 vectors during training: RAFT+AV1. Note that we use MPI Sintel dataset for model refinement because of the similarities with our test dataset, Spring dataset~\cite{spring} which is also a 3D rendered dataset from Blender.



\subsection{Investigating the Impact of Encoder Speed Presets in AV1}

The quality of the initial motion vectors is heavily influenced by the encoder configuration. The \texttt{cpu-used} parameter in the libaom-av1 encoder controls a trade-off between encoding speed and quality by adjusting the complexity of the motion search. There are two Encoder setting ranges, Slow (cpu-used between 0 and 6) and Fast (cpu-used between 7 and 12). Different motion estimation strategies are employed for each cpu-used setting.
For example, in \texttt{cpu-used=1}), the encoder performs an exhaustive, hierarchical search for motion vectors, exploring the full range of sub-pixel (pel$\in \{1, \frac{1}{2}, \frac{1}{4}, \frac{1}{8} \}$ refinements to maximise accuracy. While in Fast Encoder setting (higher cpu-used values) modes the encoder employs aggressive pruning of the search space.


\section{Experimental Setup}
    

\begin{figure*}
    \centering
    \begin{tabular}{cccc}
      \includegraphics[width=0.22\linewidth]{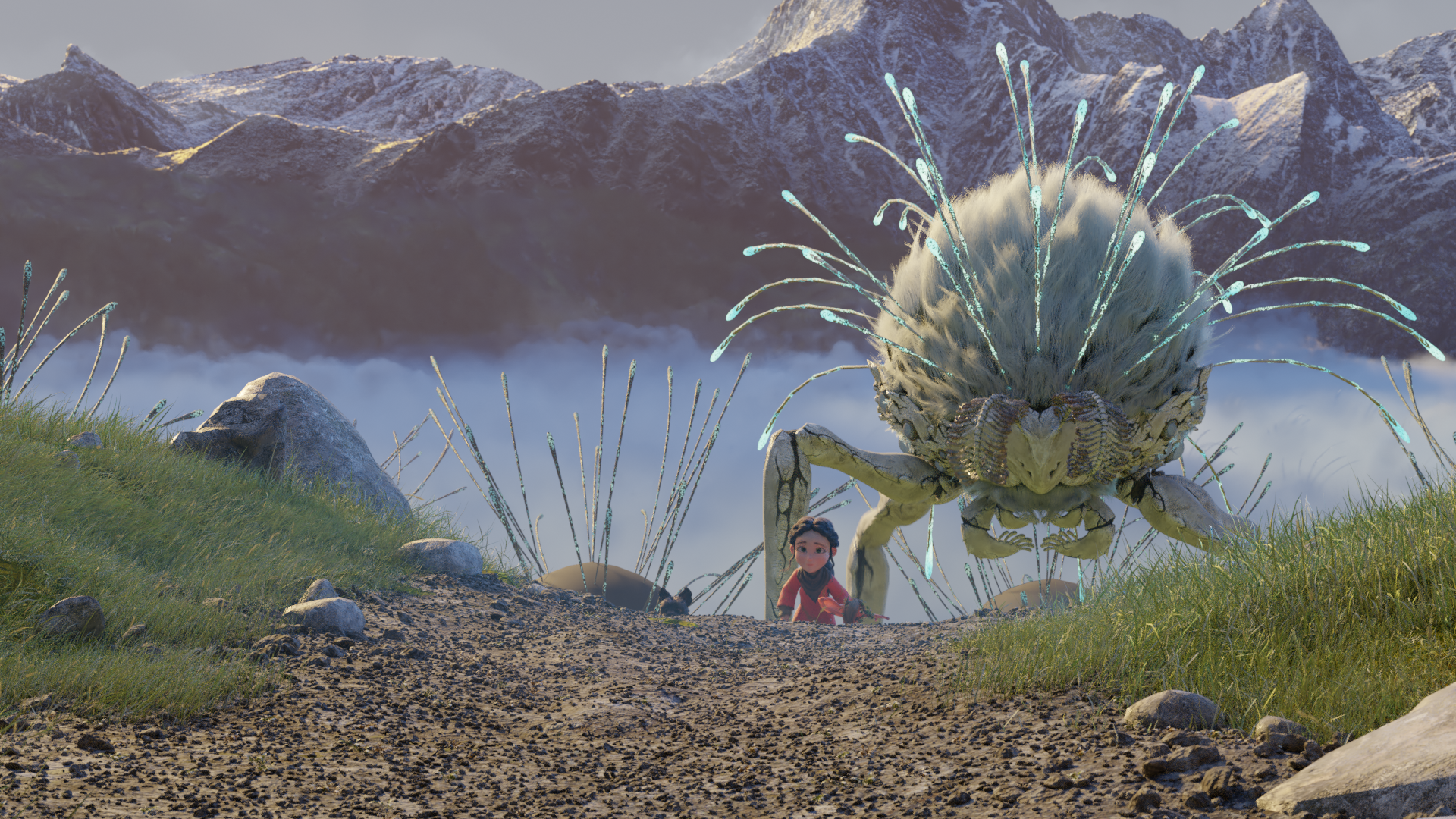}   & 
      \includegraphics[width=0.22\linewidth]{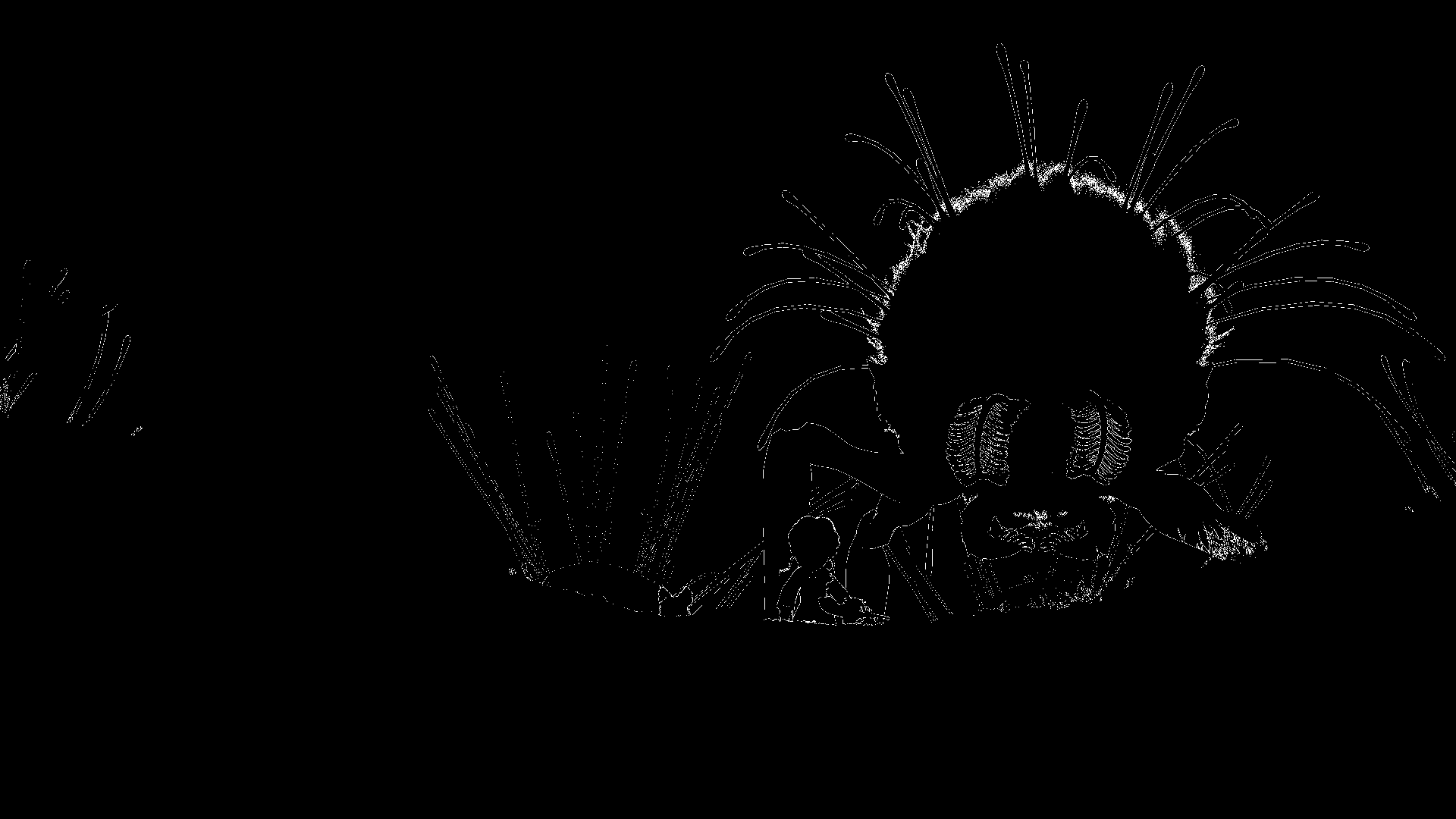} & 
      \includegraphics[width=0.22\linewidth]{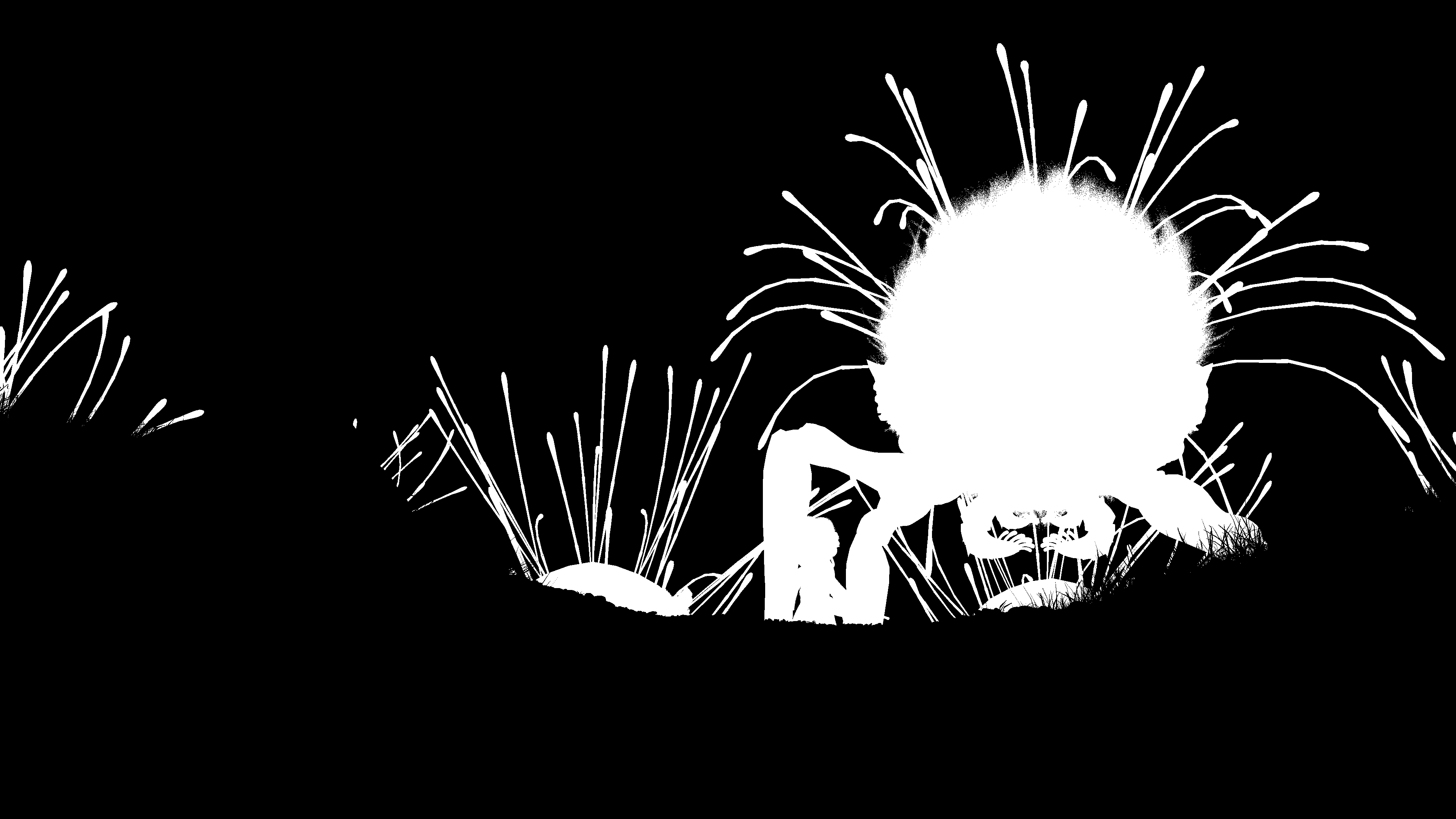} &
      \includegraphics[width=0.22\linewidth]{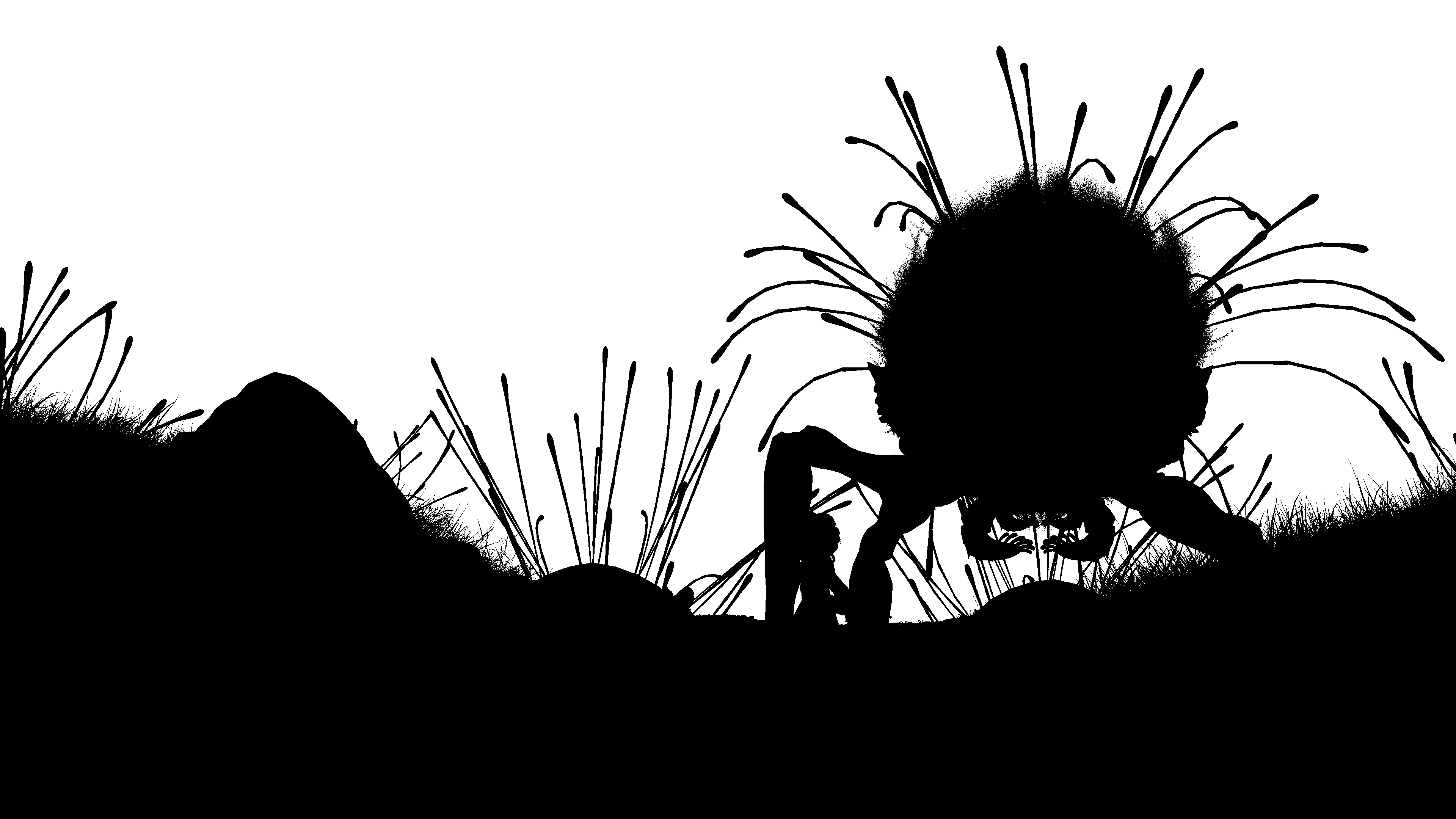}
    \end{tabular}
    \caption{Visualisation of the elements from Spring Dataset. Left to Right: Frame extracted from Sequence 38, {\em High-detail} map with white pixels corresponding to details, {\em Rigid Map} where Non-Rigid object is in white; {\em Sky Map:} where white pixels correspond to the sky.}
    \label{spring_map_example}
\end{figure*}
        
Evaluation was performed using the Spring Dataset~\cite{spring}, which comprises sequences extracted from the Blender-produced Spring animation film. 
We use the ``training split" dataset subcategory, consisting of 37 sequences at 1080p accompanied by ground truth optical flow, generated within Blender.
Semantic maps indicating {\em high-detail}, {\em rigidity} and {\em sky}, are provided with the dataset. We use these for a deeper analysis of motion quality within various scene regions.
Rigidity maps segment scenes into areas where motion is induced only by the camera and areas where objects move independently of the camera (Non-Rigid Objects).
Figure~\ref{spring_map_example} illustrates a representative frame and its associated semantic maps.
We also analyse the performances regarding the motion magnitude, having short motion for vectors with a magnitude lower than 10, Medium Motion for vectors with a magnitude greater or equal to 10 and lower than 40 and Big Motion for vectors with a magnitude greater or equal to 40.

\subsection{Encoder parameters}

For AV1 encoding, we employed libaom-av1 (\texttt{v3.12.1}) as the software encoder, and FFMPEG (\texttt{n.6.0-22}) to access the NVIDIA AV1 hardware implementation (nvenc).
Motion vectors from AV1-encoded IVF files were extracted using the \texttt{inspect} tool from the encoder's library.
HEVC encoding was performed using the H.M. (18.0) reference software, and a fork of libde265 to decode and extract motion vectors.

The HEVC encoder configuration followed the methodology of Taubman~\cite{hevc-epic}, utilising default Random Access settings, and defining GOP size of two frames (\textit{IBPB} structure) and a QP offset of 2 for B-frames, motion search range was set to 64 pixels, with $\frac{1}{4}$ precision for motion estimation precision. 
For AV1, we tried to follow the same encoder configuration. 
All motion-related parameters were left at their defaults, except the \texttt{cpu-used} parameter, which was varied for experimental analysis. The Constant Quality (CQ) parameter was set to 27 for AV1, and 32 for HEVC, after an experiment selecting the best CQ per encoder. CQ ensures that every frame gets the number of bits it deserves to achieve a certain quality level.

\section{Experimental Results and Discussion}

To evaluate the integrity of estimated motion fields we employ endpoint error (EPE) as well as Motion Compensated MSE (MCMSE) with reference to ground truth provided by Spring Dataset. Additionally, we utilise the provided semantic maps to gain deeper insights into the quality of the motion vectors. We introduce a {\em coverage} metric, representing the percentage of pixels which express both a non-zero ground truth vector and a non-zero bitstream vector. In conjunction with the semantic maps, this allows us to diagnose problems. We compute the VMAF~\cite{vmaf_paper} to verify the quality of the encoded video. This allows us to confirm that the extracted motion would be usable for processing the very same images which are being compressed.

\subsection{Impact of encoder configuration}

We compare six configurations: AOM AV1 with cpu-used 1 (libaom-1), cpu-used 6 (libaom-6), cpu-used 7 (libaom-7), and cpu-used 10 (libaom-10), alongside NVIDIA AV1 (NVENC-AV1) and HEVC reference software (hevc-hm).

\begin{figure}[t]
    \centering
    \includegraphics[width=0.8\linewidth]{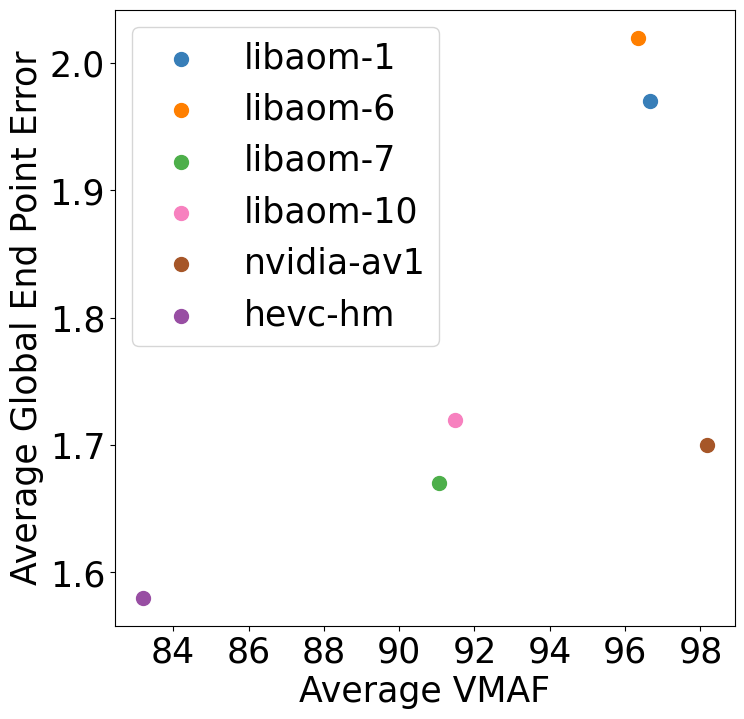}
    \caption{Average End Point Error versus Average VMAF by CPU configuration.}
    \label{fig:epe_x_vmaf}
\end{figure}

Table~\ref{table-combined-results} shows that libaom-6 performs better than other configurations in most categories, however, performance drops when we look at Medium Motion and Big Motion having respectively an EPE of +2.27 pixels and +6.98 pixels compared to HEVC correlating with a low coverage (73.82\% for Medium Motion and 31.57\% for Big Motion). AV1 ensuring better quality Figure~\ref{fig:epe_x_vmaf}, and, Medium and Big Motion yielding an important motion blur, the motion vector coverage when using AV1 is lower, yielding a higher end-point error. Libaom-1 and libaom-6 outperforms other solutions regarding Sky motion, having better coverage and Average EPE.
We can also see that NVENC-AV1 is always close to best performances having a maximum Median EPE difference lower than 0.3 pixels, except for Big Motion (+1.52 pixels).

In terms of global average EPE across all frames and sequences, the observed variation between the best and worst configurations is only 0.44 pixels, with hevc-hm achieving the lowest average EPE and libaom-6 the highest. While libaom-1 initially appears less accurate than hevc-hm, the median EPE tells a different story: libaom-1 achieves a median EPE of 0.36 pixels, outperforming the 0.43 pixel median reported for hevc-hm. These results can be explained by noting that the mean and median statistic yield very different absolute scales. This shows that outliers are significantly affecting evaluation.


Figure~\ref{fig:epe_x_vmaf} presents a comparison between average EPE and average VMAF, where optimal configurations exhibit both low EPE and high VMAF. Although the EPE difference across all configurations remains below 0.5 pixels, only the HEVC configuration (hevc-hm) falls below a VMAF of 85. NVIDIA's encoder, libaom-1, and libaom-6 all exceed VMAF 95, suggesting they deliver both high quality video output and high quality motion fields.

MCMSE results further support the quality of extracted motion fields. Motion fields generated from encoded bitstreams yield a median MCMSE that is lower than that achieved using ground truth optical flow. Specifically, ground truth flow yields a median MCMSE of 16.50, while hevc-hm achieves 12.78 and libaom-1 reaches 14.51. Notably, the NVIDIA AV1 encoder produces the best result with a MCMSE of 12.43.

Finally, regarding motion vector coverage, libaom-1 produces the most complete field among all configurations tested, covering on average 81.31\% of ground truth-moving pixels. This surpasses the 73.70\% coverage observed for the hevc-hm baseline and means that AV1 encoders seem to be better at capturing motion in the first instance with almost 9\% more coverage for libaom-1 and libaom-6 compared to hevc-hm. NVENC AV1 having less coverage than other encoders.

\begin{table*}
\caption{End Point Error (EPE) and Motion Vector (MV) Coverage (\%) over all sequences and frames from the Spring Dataset. Values are presented as Median (Average). For each metric row, the \textbf{bold value} is the best overall result, and the \underline{underlined value} is the best among the AV1 codecs (\texttt{cpu-used} $\in$ 1, 6, 7, 10).}
\label{table-combined-results}
\centering
\begin{tabular}{@{}llcccccc@{}}
\toprule
\multirow{2}{*}{Category} & \multirow{2}{*}{Metric} & \multicolumn{4}{c}{libaom-av1 \texttt{cpu-used}} & \multicolumn{1}{l}{\multirow{2}{*}{NVENC AV1}} & \multicolumn{1}{l}{\multirow{2}{*}{HEVC-HM}} \\ \cmidrule(lr){3-6}
 &  & 1 & 6 & 7 & 10 & \multicolumn{1}{l}{} & \multicolumn{1}{l}{} \\ \midrule
\multirow{2}{*}{Global} & EPE & 0.36 (1.97) & {\ul \textbf{0.34 (2.02)}} & 0.40 (1.67) & 0.44 (1.72) & 0.37 (1.70) & 0.43 (1.58) \\
 & MV Cov. (\%) & 92.63 (81.31) & {\ul \textbf{92.71 (80.20)}} & 85.39 (74.71) & 81.68 (72.15) & 82.25 (73.12) & 84.01 (73.20) \\ \midrule
\multirow{2}{*}{High Details} & EPE & 0.95 (4.85) & {\ul \textbf{0.86 (4.82)}} & 0.90 (4.05) & 0.92 (4.18) & 0.89 (4.33) & 0.91 (4.31) \\
 & MV Cov. (\%) & 84.59 (75.04) & 85.86 (75.08) & {\ul 94.33 (87.49)} & 93.65 (86.87) & 88.68 (80.97) & \textbf{95.08 (89.32)} \\ \midrule
\multirow{2}{*}{Low Details} & EPE & 0.35 (1.93) & {\ul \textbf{0.34 (1.98)}} & 0.39 (1.64) & 0.43 (1.69) & 0.36 (1.67) & 0.42 (1.54) \\
 & MV Cov. (\%) & 92.69 (80.60) & {\ul \textbf{92.80 (79.53)}} & 85.33 (73.94) & 81.68 (71.40) & 82.26 (72.34) & 83.99 (72.84) \\ \midrule
\multirow{2}{*}{Rigid Objects} & EPE & 0.22 (1.48) & {\ul \textbf{0.21 (1.53)}} & 0.30 (1.27) & 0.32 (1.31) & 0.27 (1.24) & 0.33 (1.14) \\
 & MV Cov. (\%) & 93.43 (78.70) & {\ul \textbf{93.56 (77.59)}} & 85.09 (70.73) & 81.20 (68.09) & 82.14 (69.58) & 83.36 (69.13) \\ \midrule
\multirow{2}{*}{Non-Rigid Objects} & EPE & 1.19 (7.59) & 1.20 (7.71) & 1.14 (6.63) & 1.18 (6.73) & {\ul \textbf{1.10 (7.04)}} & 1.17 (6.94) \\
 & MV Cov. (\%) & 92.06 (81.30) & 92.38 (80.61) & {\ul 95.55 (87.23)} & 94.73 (86.10) & 91.24 (81.82) & \textbf{96.06 (88.87)} \\ \midrule
\multirow{2}{*}{Sky} & EPE & {\ul \textbf{0.42 (1.43)}} & 0.42 (1.55) & 0.55 (1.68) & 0.57 (1.82) & 0.53 (1.64) & 0.62 (1.75) \\
 & MV Cov. (\%) & {\ul \textbf{96.18 (75.51)}} & 94.53 (72.70) & 83.28 (63.23) & 69.57 (58.08) & 71.51 (61.91) & 76.39 (57.42) \\ \midrule
\multirow{2}{*}{Non-sky} & EPE & 0.36 (2.06) & {\ul \textbf{0.34 (2.10)}} & 0.39 (1.70) & 0.42 (1.76) & 0.36 (1.75) & 0.41 (1.58) \\ 
 & MV Cov. (\%) & {\ul \textbf{94.25 (81.46)}} & 93.92 (80.48) & 88.91 (76.50) & 85.88 (74.32) & 86.15 (75.56) & 88.08 (75.38) \\ \midrule \midrule
\multirow{2}{*}{Short Motion} & EPE & 0.26 (0.58) & {\ul \textbf{0.25 (0.61)}} & 0.33 (0.64) & 0.36 (0.67) & 0.30 (0.62) & 0.37(0.83) \\
 & MV Cov. (\%) & 93.55 (80.51) & {\ul \textbf{93.65 (79.37)}} & 85.48 (73.43) & 81.58 (70.91) & 82.26 (71.91) & 83.31 (71.84) \\ \midrule
\multirow{2}{*}{Medium Motion} & EPE & 7.64 (9.81) & 8.69 (10.42) & {\ul \textbf{5.57 (8.00)}} & 5.82 (8.11) & 6.87 (9.07) & 6.42 (8.78) \\ 
 & MV Cov. (\%) & 79.63 (67.88) & 73.82 (65.01) & {\ul 92.47 (85.35)} & 91.93 (85.45) & 82.86 (74.43) & \textbf{98.32 (90.19)} \\ \midrule
\multirow{2}{*}{Big Motion} & EPE & 52.71 (60.11) & 53.15 (60.78) & {\ul 46.47 (54.77)} & 47.35 (55.32) & 47.69 (56.13) & \textbf{46.17 (54.18)} \\
 & MV Cov. (\%) & 33.06 (39.73) & 31.57 (41.52) & {\ul 64.05 (61.75)} & 63.17 (61.37) & 51.55 (53.58) & \textbf{94.20 (78.07)} \\ \bottomrule
\end{tabular}%
\end{table*}

\subsection{Refinememt with RAFT}

\begin{figure}
    \centering
    \begin{tabular}{cc}
      \includegraphics[width=0.44\linewidth]{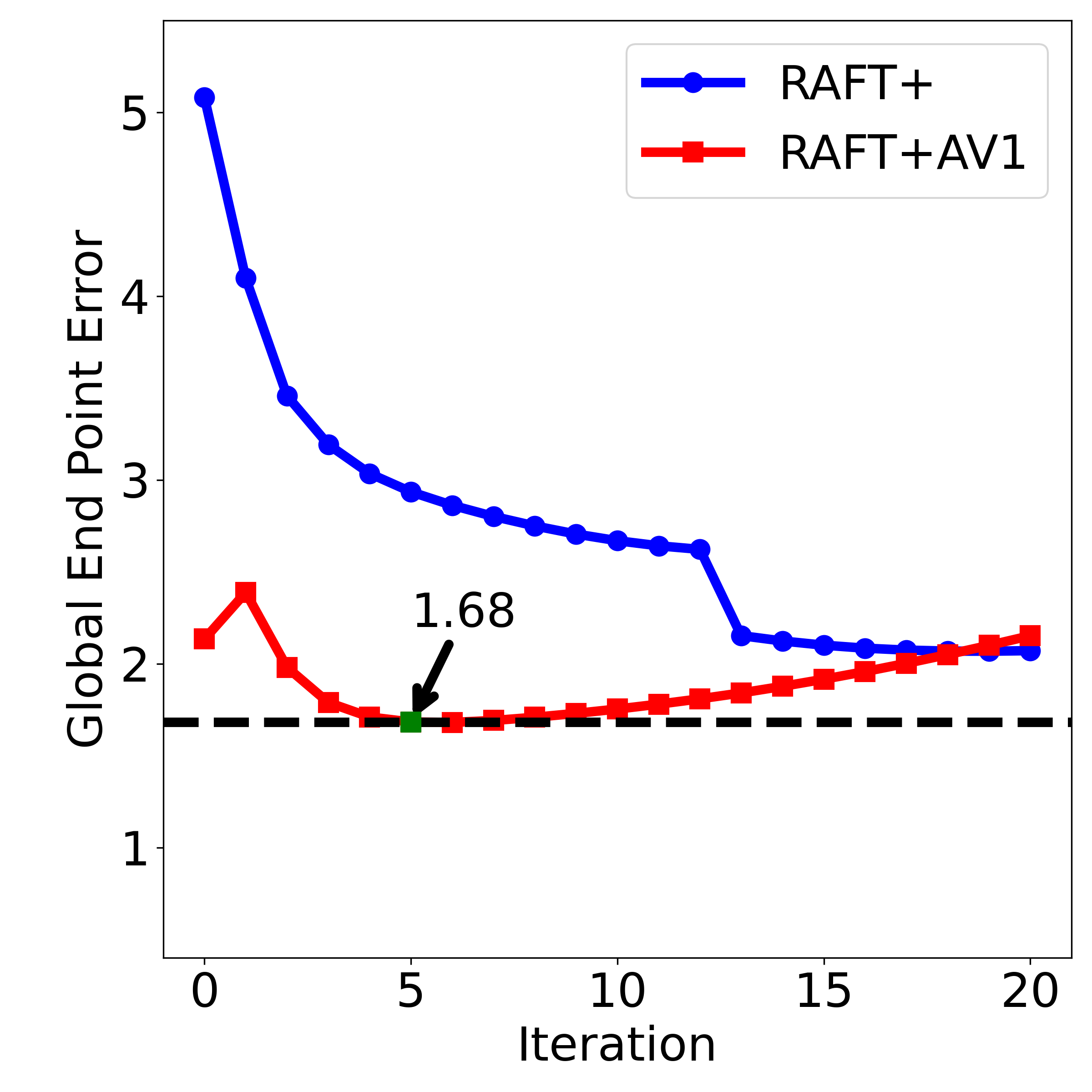}\label{kitti-epe}   & 
      \includegraphics[width=0.44\linewidth]{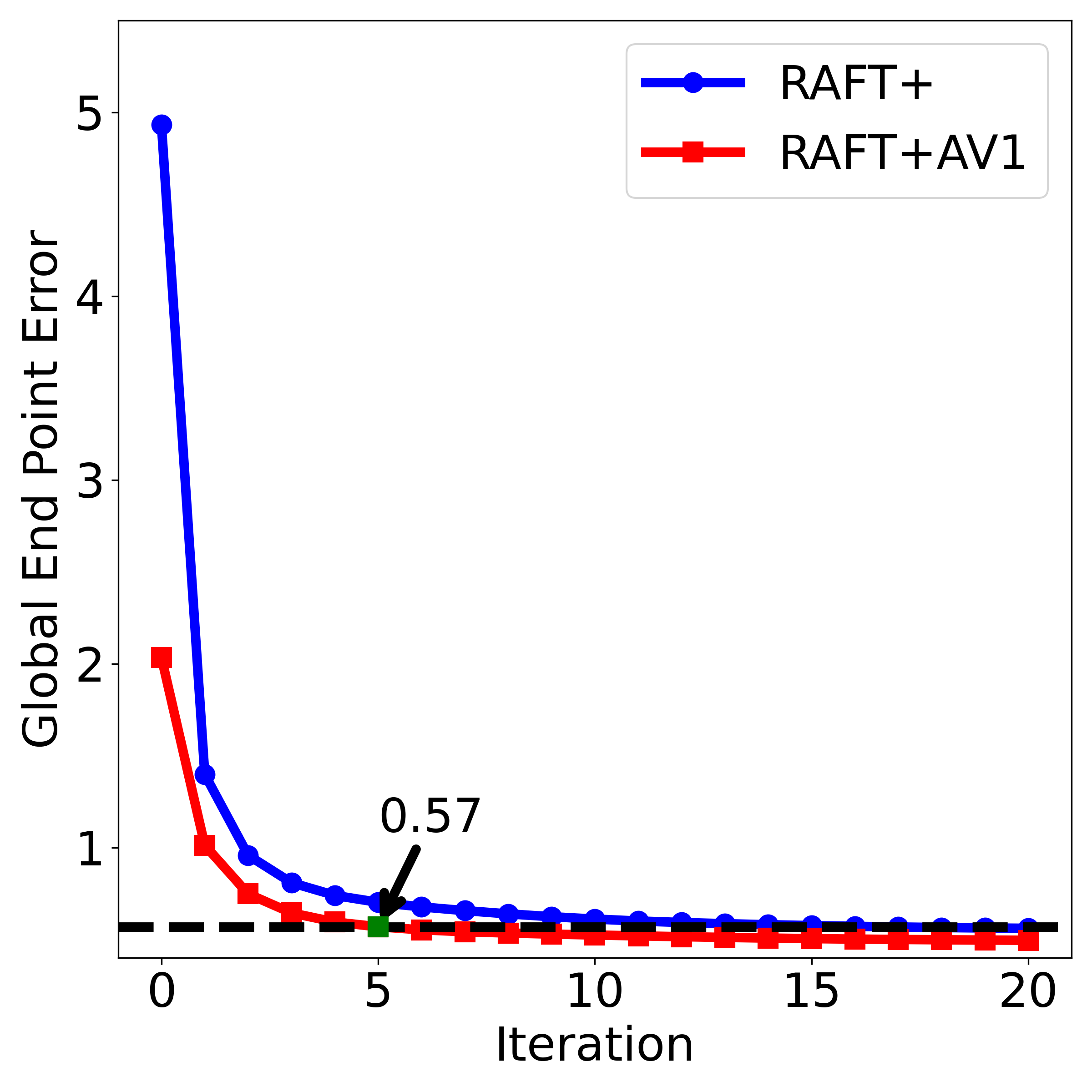} \\
      (a) & (b) \\
    \end{tabular}
    \caption{Impact of using AV1 Motion Field as warm-start for Raft. Left: Global EPE of RAFT off the shelf; Right: Global EPE of Fine-tuned RAFT; Using AV1 vectors as warm start (red) does not affect final accuracy that much but always leads to better convergence regardless of RAFT refinement. 
    }
    \label{RAFT-performances}
\end{figure}

As shown in Figure~\ref{RAFT-performances}-a, using AV1 motion vectors as a warm start for RAFT significantly improves the average end-point error (EPE). After only five iterations, RAFT with AV1 initialisation achieves an EPE of 1.68 pixels, already outperforming vanilla RAFT, which requires twenty iterations to reach an EPE of 2.07 pixels. We further compare RAFT fine-tuned on the Sintel dataset using AV1 initialisation against a baseline RAFT fine-tuned without warm starts. While both models ultimately converge to similar performance, warm-starting RAFT from AV1 motion vectors leads to faster convergence. As shown in Figure~\ref{RAFT-performances} (b), the model reaches an EPE of 0.57 after just five iterations, nearly matching the EPE of 0.56 achieved by the baseline after twenty iterations. 

These results suggest that AV1 motion vectors serve as effective initial estimates for deep optical flow models, enabling faster convergence and improved performance, particularly in regions of fine visual detail.



\section{Conclusion}

This study shows that motion vectors extracted from AV1 bitstreams can deliver both high video quality and accurate motion information, which is required for many practical applications. In particular, libaom provides an appealing alternative to HEVC, achieving similar motion accuracy alongside higher perceptual quality (VMAF).
The AV1 encoder’s low motion-compensated errors make these motion fields suitable for tasks such as pixel enhancement and frame interpolation, especially in real-time use cases. Furthermore, refining these sparse vectors with RAFT enables efficient, dense optical flow estimation, reducing computation compared to starting from scratch. Using AV1 motion vectors as an initialisation for deep learning models thus offers a promising route to quickly improve flow quality, particularly in high-detail regions. Future work should explore the impact of encoder's parameter on the quality of the generated motion vectors and extend this work to new encoders and real-world data. New refinement strategies will be tested. Work will be done on applications that may benefit from the extracted bitstream, like action recognition and classification or 3D reconstruction.

\bibliographystyle{IEEEtran}
\bibliography{references}

\end{document}